\documentclass{ws-p8-50x6-00}

\def \beq {\begin{equation}}
\def \eeq {\end{equation}}

\def \onbb {$0\nu\beta\beta$ }

\begin{document}
%
\title{Evidence for Neutrinoless Double Beta Decay}
\author{H.V. Klapdor-Kleingrothaus$^{1,3}$, \hspace{12.cm}
	A. Dietz$^{1}$, H.L. Harney$^{1}$, I.V. Krivosheina$^{1,2}$}

\address{$^{1}$Max-Planck-Institut f\"ur Kernphysik,
 	Postfach 10 39 80, D-69029 Heidelberg, Germany\\
	$^{2}$Radiophysical-Research Institute, Nishnii-Novgorod, Russia\\
	$^{3}$Spokesman of the GENIUS and HEIDELBERG-MOSCOW Collaborations,\\ 
	e-mail: klapdor@gustav.mpi-hd.mpg.de,\\
	 home page: http://www.mpi-hd.mpg.de/non$\_$acc/}

\maketitle              

\abstracts{The data of the HEIDELBERG-MOSCOW double beta decay
	experiment for the measuring period August 1990 - May 2000 
	(54.9813\,kg\,y or 723.44\,molyears), published recently,  
	are analyzed 
	using the potential of the Bayesian method 
	for low counting rates. 
	First evidence for neutrinoless double beta decay is observed
	giving first evidence for lepton number violation.  
	The evidence for this decay mode is 97$\%$ (2.2$\sigma$)  
	with the Bayesian method, and 99.8$\%$ c.l. (3.1$\sigma$) 
	with the method recommended by the Par\-ticle Data Group. 
	The half-life of the process is found 
	with the Bayesian method to be 
	${\rm T}_{1/2}^{0\nu} = (0.8 - 18.3) \times 10^{25} 
	{\rm~ y}$ (95\% c.l.) with a best value of 
	$1.5 \times 10^{25} {\rm~ y}$.    
	The deduced value of the effective neutrino mass is,  
	with the nuclear matrix elements from 
\cite{Sta90},
	$\langle m \rangle$ =  
	(0.11 - 0.56)\,eV (95$\%$ c.l.), 
	with a best value of  0.39\,eV. 
	Uncertainties in the nuclear matrix elements may widen  
	the range given for the effective neutrino mass 
	by at most a factor 2.
	Our observation which at the same time means evidence 
	that the neutrino is a Majorana particle, will be 
	of fundamental importance for neutrino physics.
	\hspace{5.cm}{\bf PACS.} 14.69.Pq Neutrino mass and mixing - 
	23.40.Bw Weak-interaction and lepton (including neutrino) aspects - 
	23.40.-s Beta decay; double beta decay; electron and muon capture.}


	The neutrino oscillation
	interpretation of the atmospheric and solar neutrino data,  
	deliver a strong indication for a non-vanishing neutrino mass.
	While such kind of experiments yields information on the difference of 
	squared neutrino mass eigenvalues and on mixing angles, the 
	absolute scale of the neutrino mass is still unknown.
	Information from double beta decay experiments is indispensable to
	solve these questions 
\cite{KKPS1,KK60Y}.
	Another important problem is that of the fundamental
	character of the neutrino, whether it is a Dirac or a Majorana
	particle
\cite{Majorana37,Rac37}.  
	Neutrinoless double beta decay could answer also this question. 
	Perhaps the main question, which can be investigated 
	by double beta decay with high sensitivity, is  
	that of lepton number conservation or non-conservation.

	Double beta decay, the rarest known nuclear decay process, can occur
	in different modes:

\begin{tabbing}
\hspace*{4.5cm} \= \hspace*{7.9cm} \= \kill
\qquad $ 2\nu\beta\beta- \rm decay:$ \> $ A(Z,N) \rightarrow
A(Z\!+\!2,N\!-\!2) + 2e^- + 2\bar{\nu}_e$ \hspace*{1.05cm}(1)\\[-0ex]
\qquad $ 0\nu\beta\beta- \rm decay:$ \> $ A(Z,N) \rightarrow
A(Z\!+\!2,N\!-\!2) + 2e^-$ \hspace*{2cm}(2) \\[-0ex]
\qquad $ 0\nu(2)\chi\beta\beta- \rm decay:$ \> $ A(Z,N) \rightarrow
A(Z\!+\!2,N\!-\!2) + 2e^- + (2)\chi$ \hspace*{0.9cm}(3) \\[-0ex]
\end{tabbing}

	While the two-neutrino mode (1)
	is allowed by the Standard Model of particle physics, 
	the neutrinoless mode 
	(\onbb) (2) requires violation of lepton number 
	($\Delta$L=2). 
	This mode is possible only, if the neutrino 
	is a Majorana particle, i.e. the neutrino is its own antiparticle 
	(E. Majorana
\cite{Majorana37},  
	G. Racah 
\cite{Rac37},    
	 for subsequent works we refer to 
\cite{Mcl57,Case57,Ahl96}, 
	 for some reviews see   	
\cite{Doi85,Mut88,KK60Y,Vog2001}). 
	First calculations of \onbb decay based 
	on the Majorana theory have been done by W.H. Furry 
\cite{Fur39}.

	Neutrinoless double beta decay can not only probe a
	Majorana neutrino mass, but various new physics scenarios beyond the
	Standard Model, such as R-parity violating supersymmetric models, 
	R-parity conserving SUSY models,  
	leptoquarks, 
	violation of Lorentz-invariance,  
	and compositeness 
	(for a review see 
	\cite{KK60Y,KK-LeptBar98,KK-SprTracts00}). 
	Any theory containing lepton number violating
	interactions can in principle lead to this process allowing to obtain
	information on the specific underlying theory.
	The experimental signature of the neutrinoless mode is a peak at the
	Q-value of the decay.

	The HEIDELBERG-MOSCOW double beta decay experiment in the Gran Sasso
	Underground Laboratory searches
	for double beta decay of 
	$^{76}{Ge} \longrightarrow ^{76}{Se}$ + 2 $e^-$ + (2$\nu$) 
	since 1990. 
	It is the most sensitive double beta experiment since 
	almost eight years now.	
 	The experiment operates five enriched (to 86$\%$) high-purity 
	$^{76}{Ge}$ detectors, with a total mass of 11.5\,kg,  
	the active mass of 10.96\,kg being equivalent to a source 
	strength of 125.5 mol $^{76}{Ge}$ nuclei. This  
	is the largest source strength in use. 

	The high energy resolution of the Ge detectors assures that there 
	is no background for a \onbb line from the two-neutrino 
	double beta decay in this experiment.

	The unique feature of neutrinoless double beta decay is that 
	a measured half-life allows to deduce information on 
	the effective Majorana neutrino mass
$\langle m \rangle $, 
	which is a superposition of neutrino mass eigenstates. 
	The half-life is given, when ignoring contributions 
	from right-handed weak currents, by 
\cite{Doi85,Mut88}.
	\beq 
[T^{0\nu}_{1/2}(0^+_i \rightarrow 0^+_f)]^{-1}= C_{mm} 
\frac{\langle m \rangle^2}{m_{e}^2}
	\eeq

\beq{}
\langle m \rangle = 
|m^{(1)}_{ee}| + e^{i\phi_{2}} |m_{ee}^{(2)}|
+  e^{i\phi_{3}} |m_{ee}^{(3)}|~,
\label{mee}
\eeq
	Here $|m_{ee}^{(i)}| \exp{(i \phi_i)}$ 
	($i = 1, 2, 3$)  are  the contributions to $\langle m \rangle$
	from individual mass eigenstates, 
	with  $\phi_i$ denoting relative Majorana phases connected 
	with CP violation and, 
	$C_{mm}$ denotes a nuclear matrix element, 
	which can be calculated.

	The effective mass is closely related to the parameters of neutrino
	oscillation experiments.
	The importance of $\langle m \rangle$ for solving the problem 
	of the structure 
	of the neutrino mixing matrix and in particular to fix 
	the absolute scale of the neutrino
	mass spectrum which cannot be fixed by $\nu$ - oscillation 
	experiments alone,  
	has been discussed in detail in e.g. 
\cite{KKPS1}.

	In this paper, we present a new, refined analysis of the data  
	obtained in the HEIDELBERG-MOSCOW experiment during the 
	period August 1990 - May 2000
\cite{HDM01}.
	The analysis concentrates on the 
	neutrinoless decay mode which is the one relevant for
	particle physics (see, e.g.
\cite{KKPS1,KK60Y}). 
	A description of the HEIDELBERG-MOSCOW experiment has been 
	given recently in 
\cite{HDM97}, and in 
\cite{HDM01}.  

 

\begin{figure}[t]

\vspace{-0.4cm}
\begin{center}
\includegraphics
[scale=0.33]{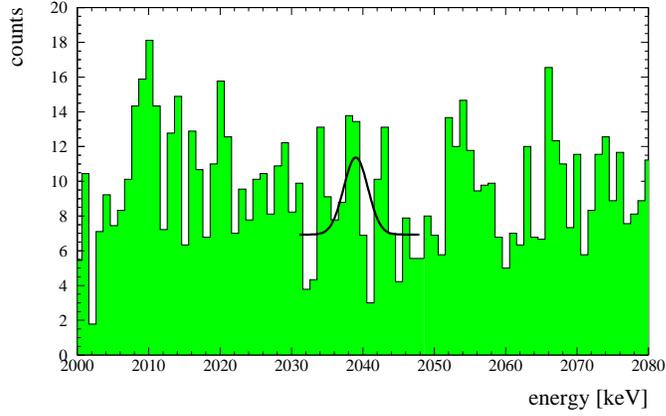} 
\end{center}

\vspace{-0.8cm}
\caption[]{Sum spectrum of the 
	$^{76}{Ge}$ detectors Nr. 1,2,3,4,5 over the period 
	August 1990 to May 2000, 
	(54.981\,kg\,y) in the energy interval 2000 - 2080\,keV, around the 
	Q$_{\beta\beta}$ value of double beta decay 
	(Q$_{\beta\beta}$~=~2039.006(50)\,keV). 
	The curve results from Bayesian inference in the 
	way explained in the text. It corresponds to a half-life 
	T$_{1/2}^{0\nu}$=$(0.80 - 35.07) \times 10^{25}$ (95\% c.l.).
}
\label{Sum_spectr_Alldet_1990-2000}
\end{figure}

\begin{figure}[t]

\vspace{-0.4cm}
\begin{center}
\includegraphics[scale=0.33]{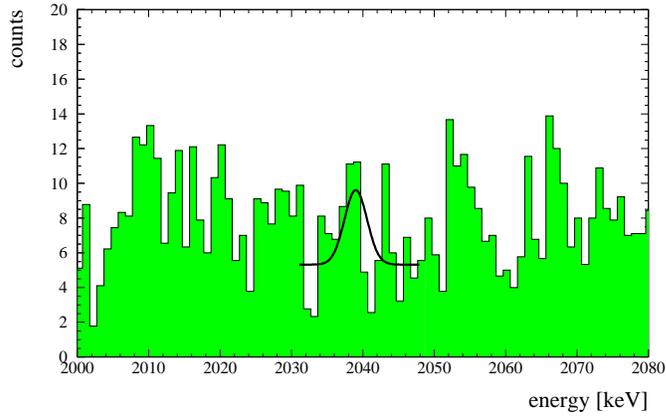} 
\end{center}

\vspace{-0.8cm}
\caption[]{Sum spectrum of the 
	$^{76}{Ge}$ detectors Nr. 1,2,3,5 over the period 
	August 1990 to May 2000, 
	46.502\,kg\,y. 
	The curve results from Bayesian inference in the 
	way explained in the text. 
	It corresponds to a half-life 
	T$_{1/2}^{0\nu}$=(0.75 - 18.33)$\times~10^{25}$~y 
	(95\% c.l.).
}
\label{Sum_spectr_1-2-3-5det_95-2000}
\end{figure}

\begin{figure}[t]

\vspace{-0.4cm}
\begin{center}
\includegraphics[scale=0.33]
{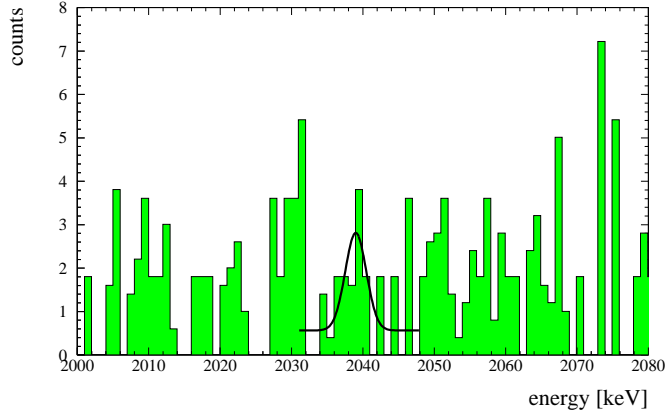} 
\end{center}

\vspace{-0.8cm}
\caption[]{Sum spectrum, 
	measured with the detectors 
	Nr. 2,3,5 operated with pulse shape analysis in the period 
	November 1995 to May 2000 (28.053\,kg\,y),  
	in the region of interest for the \onbb{}- decay. 
	Only events identified as single site events (SSE) 
	by all three pulse shape analysis methods 
\cite{HelKK00,KKMaj99}
	have been accepted.
	The spectrum has been corrected for the 
	efficiency of SSE identification (see text).
	The curve results from Bayesian inference in the 
	way explained in the text. 
	The signal corresponds to a half-life 
	T$_{1/2}^{0\nu}$=$(0.88 - 22.38) \times 10^{25}$~y (90\% c.l.).
}
\label{Sum_spectr_5det}
\end{figure}

	Fig.
\ref{Sum_spectr_Alldet_1990-2000} 
	shows the combined spectrum of 
	the five enriched detectors 
	obtained over the period 
	August 1990 - May 2000, with a statistical significance 
	of 54.981\,kg\,y (723.44\,molyears), around 
	the Q-value of double beta decay. 
	The latter has been determined to be 
	Q$_{\beta\beta}$~=~2039.006(50) keV in a recent precision 
	experiment
\cite{New-Q-2001}.
	Fig. 
\ref{Sum_spectr_1-2-3-5det_95-2000} 
	shows the spectrum obtained with detectors Nr. 1,2,3,5 over 
	the period August 1990 - May 2000, with 
	a significance of 46.502\,kg\,y.
	Fig. 
\ref{Sum_spectr_5det} shows the spectrum of single site events (SSE) 
	obtained for detectors 2,3,5 
	in the period November 1995 - May 2000, 
	under the restriction that the signal simultanously fulfills  
	the criteria of {\it all three} methods 
	of pulse shape analysis we have recently developed 
\cite{HelKK00,KKMaj99} 
	and with which we operate     all detectors except detector 1
	(significance 28.053\,kg\,y). 
	Double beta events are single site events 
	confined to a few mm region in the detector,  
	corresponding to the track length of the emitted electrons, 
	in contrast to e.g. multiple scattering $\gamma$ - events.  
	From simulation we expect that about 5$\%$  
	of the double beta single site events should be seen also 
	as MSE. 
	This is caused by bremsstrahlung of the emitted electrons
\cite{Diss-Dipl}.

	All the spectra are obtained 
	after rejecting coincidence events between different 
	Ge detectors and events coincident with activation 
	of the muon shield. 
	The spectra which are taken in bins of 0.36\,keV are shown
	in Figs. 
\ref{Sum_spectr_Alldet_1990-2000},\ref{Sum_spectr_1-2-3-5det_95-2000},\ref{Sum_spectr_5det}  
	in 1\,keV bins, which explains 
	the broken number in the ordinate. 
	We do the analysis of the measured spectra 
	with (Fig. 
\ref{Sum_spectr_Alldet_1990-2000}) 
	and without the data of detector 4 (Figs. 
\ref{Sum_spectr_1-2-3-5det_95-2000},\ref{Sum_spectr_5det}) 
	since the latter 
	does not have a muon shield and has the weakest 
	energy resolution. 
	We ignore for each detector the first 200 days of operation, 
	corresponding to about three half-lives of $^{56}{Co}$  
	($T_{1/2}$ = 77.27\,days), 
	to allow for some decay of short-lived radioactive impurities.

	The background rate in the energy range 2000 - 2080\,keV 
	is found to be (0.17 $\pm$ 0.01)\,events/\,kg\,y\,keV 
	({\it without} pulse shape analysis)  
	considering {\it all} data as background. 
	This is the lowest value obtained in such type of experiments.
	The energy resolution at the Q$_{\beta\beta}$ value 
	in the sum spectra 
	is extrapolated from the strong lines in the spectrum 
	to be (4.00 $\pm$ 0.39)\,keV for the spectra with detector 4, and 
	(3.74 $\pm$ 0.42)\,keV (FWHM) for those without detector 4.
	The energy calibration of the experiment has an uncertainty
	 of 0.20\,keV. For more experimental details see 
\cite{HDM97,HDM01}.


	We analyse the spectra shown in Figs.
\ref{Sum_spectr_Alldet_1990-2000},\ref{Sum_spectr_1-2-3-5det_95-2000},\ref{Sum_spectr_5det} 
	with the following methods:

	1. {\sf Bayesian method}, which is used widely
	at present (see, e.g. 
\cite{BayesMeth}). 
	This method is particularly suited for low
	counting rates, where the data follow 
	a Poisson distribution, that cannot be approximated by a Gaussian 
	(see, also  
\cite{Harn01-prep,Harn02-monogr}).

	2. {\sf Method recommended by the Particle Data Group} 
\cite{RPD98}.

	The ${\chi}^2$ - method 
	is not used since it is expected to have some limitation 
	for application at low counting rates, as also 
	the Maximum Likelihood Method.
	The limitation is due to the Gaussian approximations 
	inherent in these methods.

	To be conservative, we first concentrate on the  
	spectra without pulse shape analysis.
	For the evaluation, we consider the {\it raw data} of 
	the detectors. 
	After that we shall turn to the analysis of the 
	spectrum obtained with pulse shape discrimination. 

\begin{figure}[t]
\begin{center}
\includegraphics[scale=0.2]
{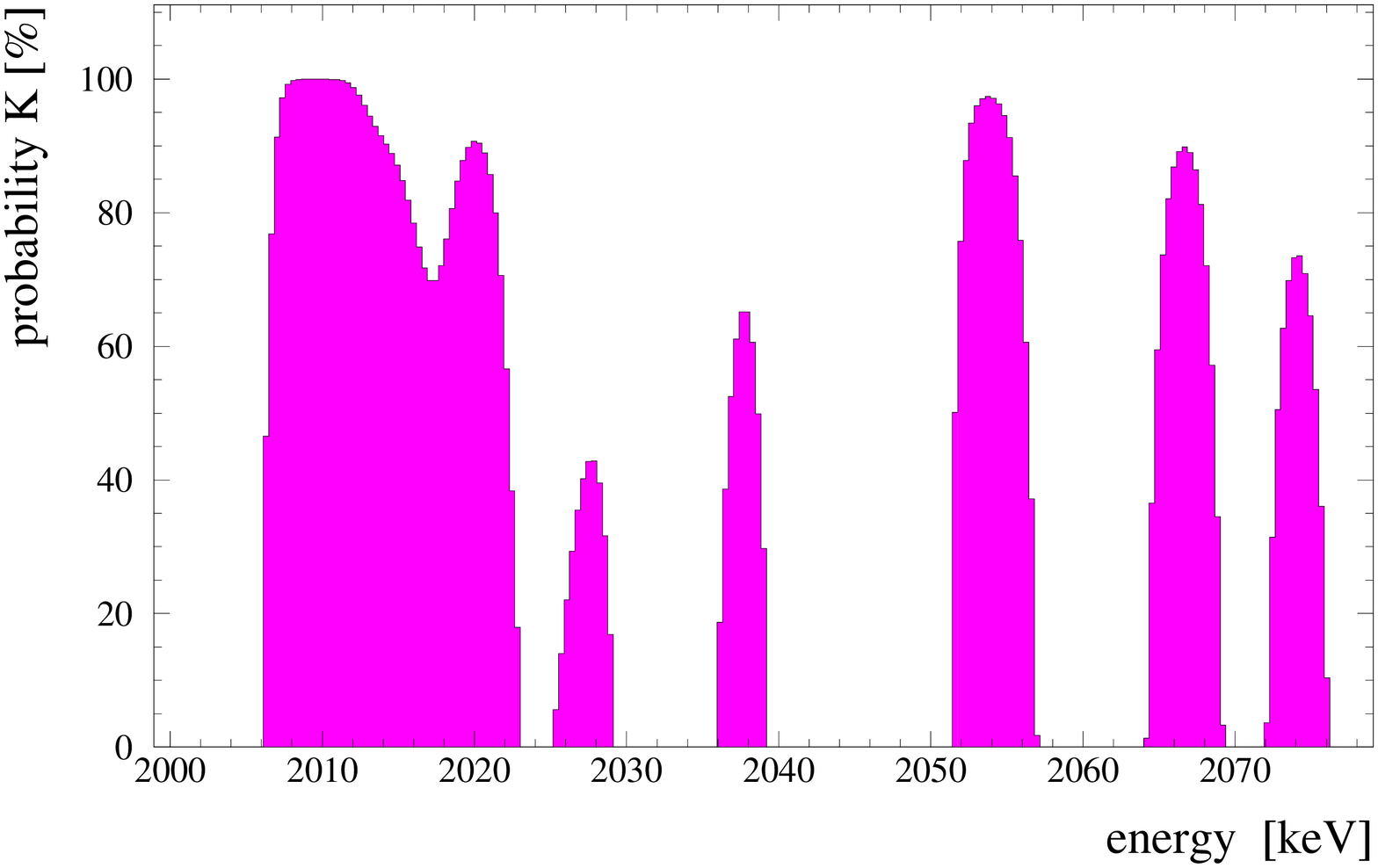} 
\includegraphics[scale=0.2]
{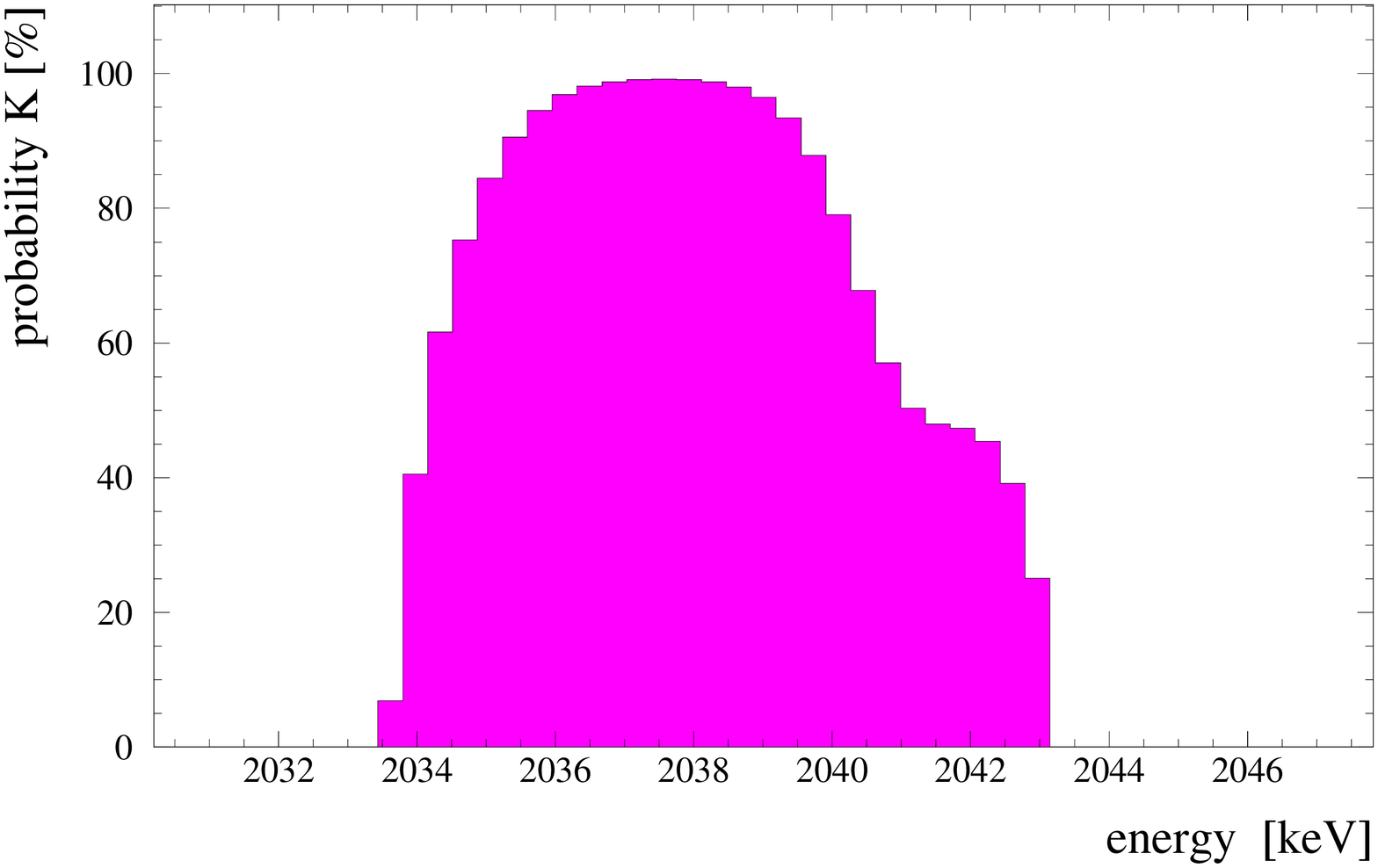} 
\end{center}

\vspace{-0.4cm}
\caption[]{Scan for lines in the full spectrum 
	taken from 1990-2000 with detectors Nr. 1,2,3,4,5, 
	(Fig. 
\ref{Sum_spectr_Alldet_1990-2000}), 
	with the Bayesian method. 
	The ordinate is the probability K that a line exists at energy E.  
	Left: Energy range 2000 -2080\,keV. 
	Right: Energy range of interest 
	around Q$_{\beta\beta}$.} 
\label{Bay-Alles-90-00-gr-kl-Bereich}
\end{figure}

\begin{figure}[t]
\begin{center}
\includegraphics[scale=0.2]
{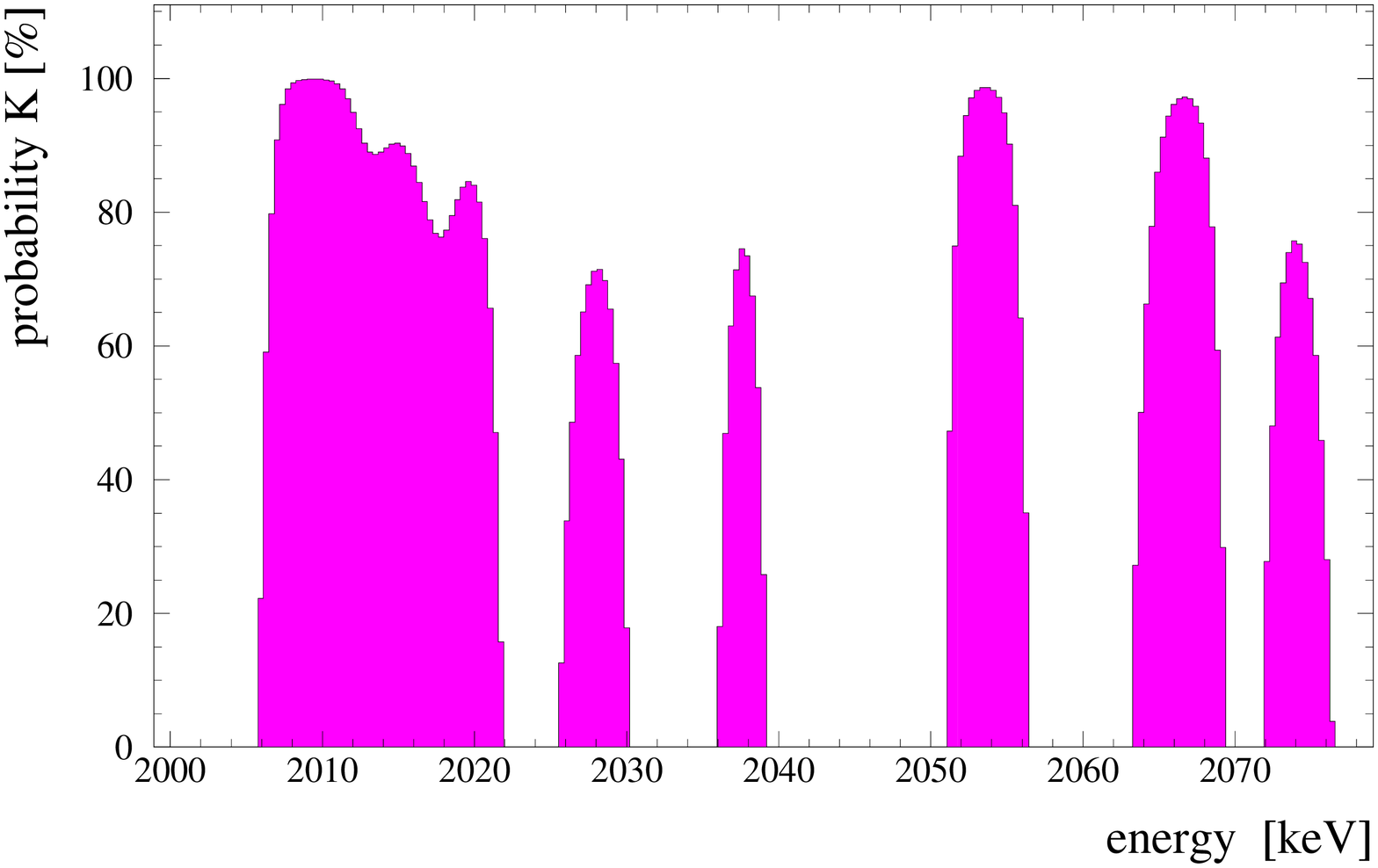}
\includegraphics[scale=0.2]
{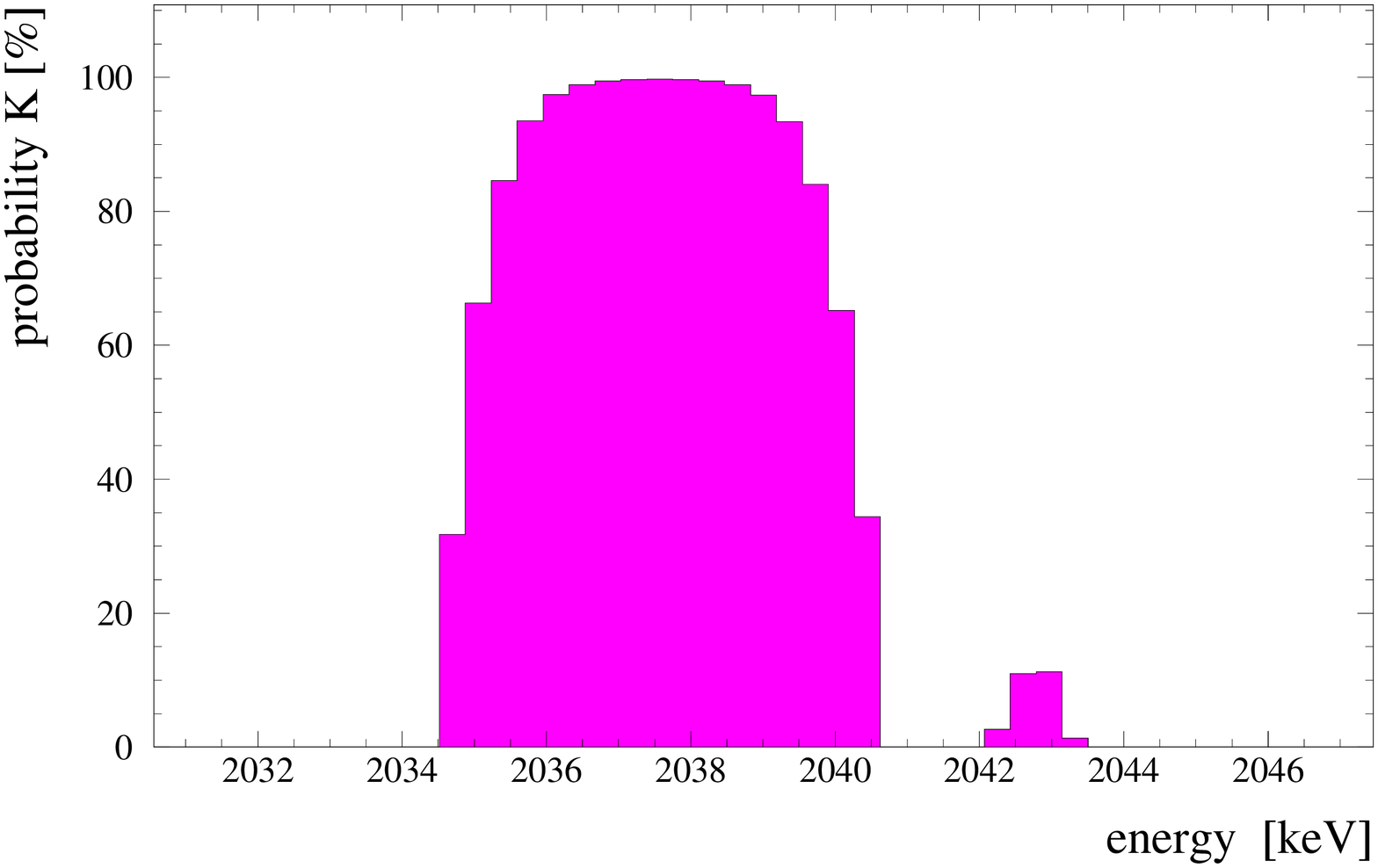} 
\end{center}
\vspace{-0.4cm}
	\caption[]{Left: 
	Probability K that a line exists at a given energy in the 
	range of 2000-2080\,keV derived via Bayesian inference 
	from the spectrum shown in Fig. 
\ref{Sum_spectr_1-2-3-5det_95-2000}.
	Right: 
	Result of a Bayesian scan for lines as in 
	the left part of this figure,  
	but in the energy range of interest around Q$_{\beta\beta}$.} 
\label{Bay-Chi-all-90-00-gr}
\end{figure}

\begin{figure}[b]
\begin{center}
\includegraphics[scale=0.2]
{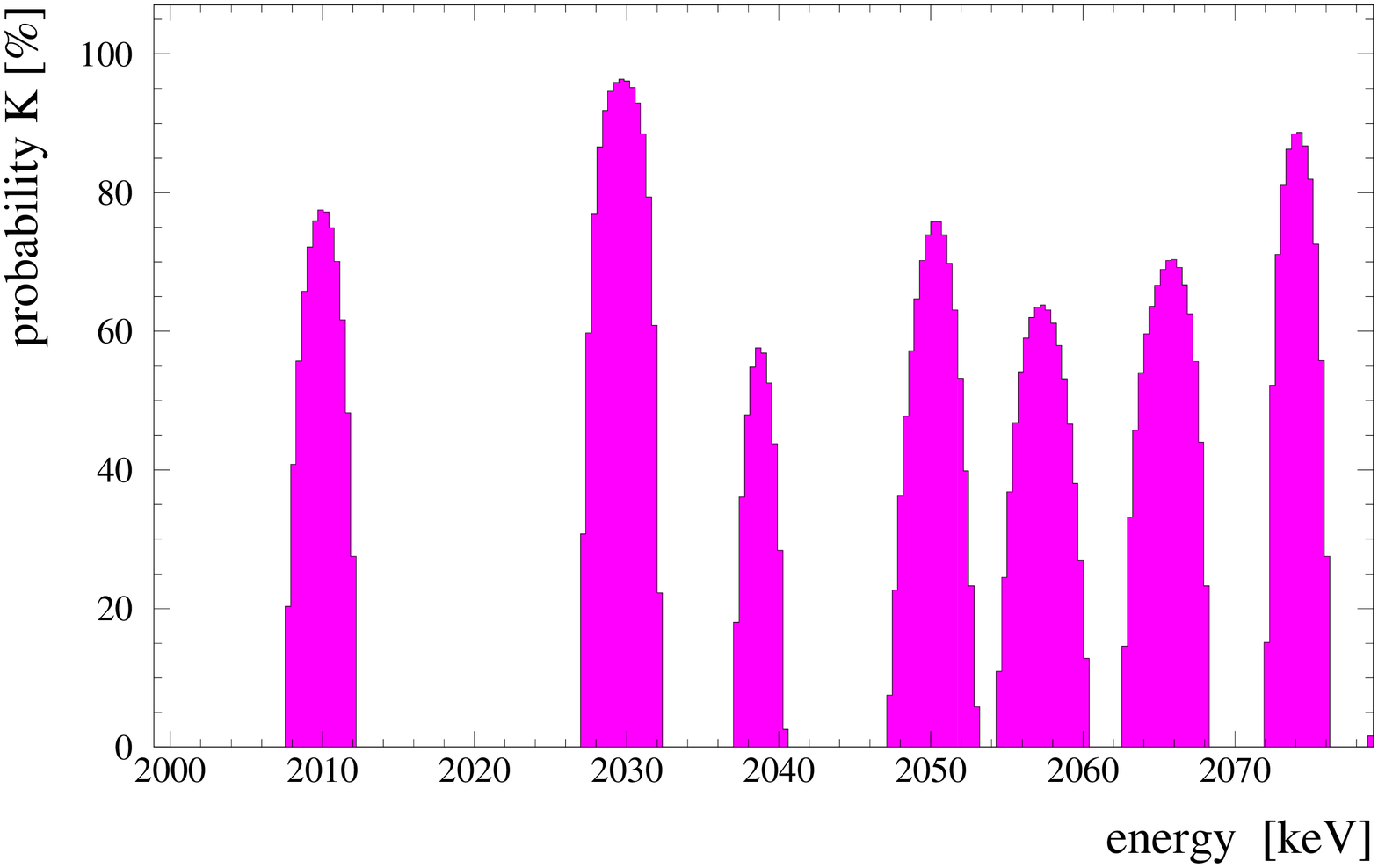} 
\includegraphics[scale=0.2]
{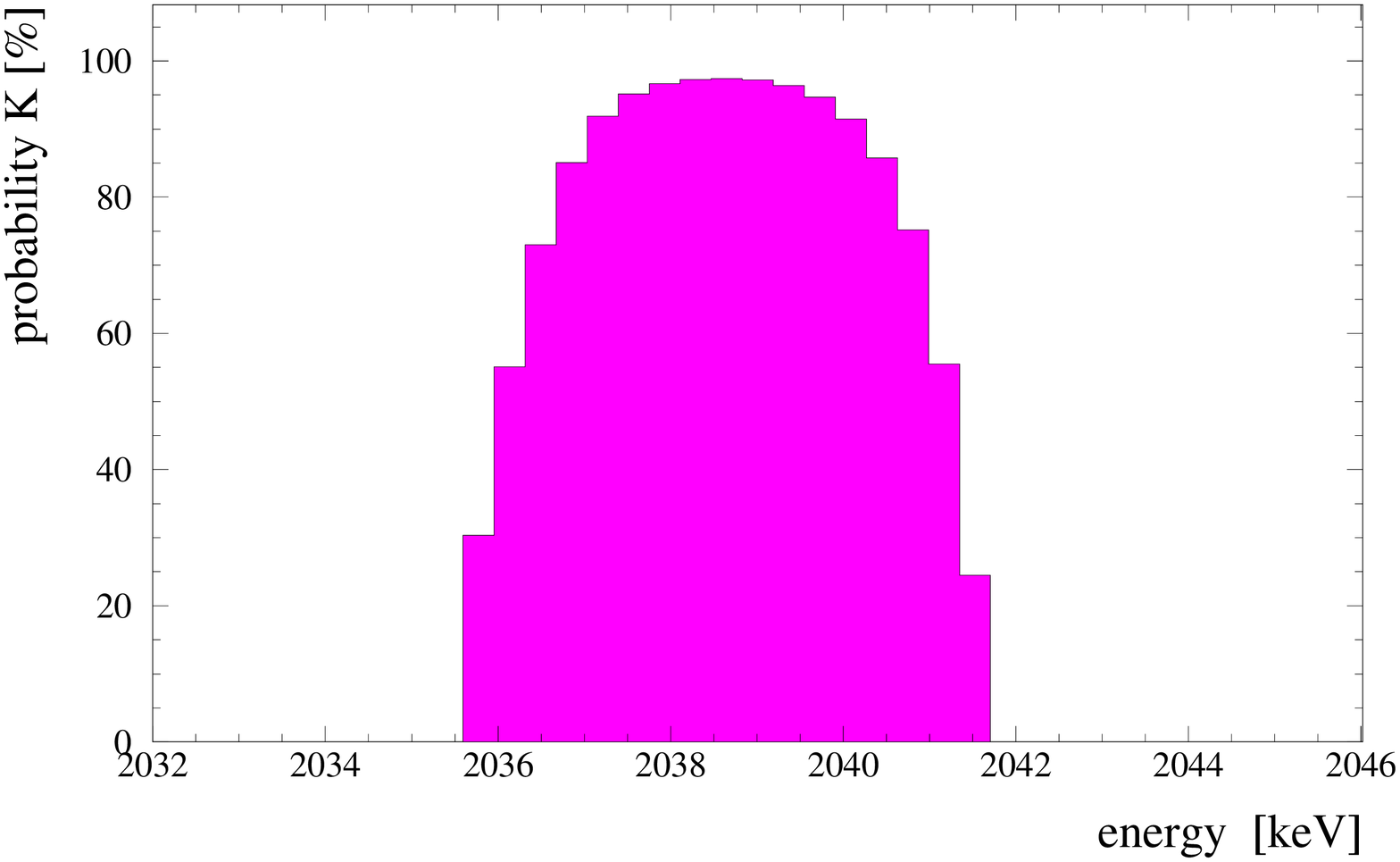}
\end{center}
\caption[]{Scan for lines in the single site event spectrum 
	taken from 1995-2000 with detectors Nr. 2,3,5, 
	(Fig. 
\ref{Sum_spectr_5det}), 
	with the Bayesian method (as in Figs.
\ref{Bay-Alles-90-00-gr-kl-Bereich},\ref{Bay-Chi-all-90-00-gr}).
	Left: Energy range 2000 -2080\,keV. 
	Right: Energy range of interest around Q$_{\beta\beta}$.} 
\label{Bay-Hell-95-00-gr-kl-Bereich}
\end{figure}

	Figs. 
\ref{Bay-Alles-90-00-gr-kl-Bereich},\ref{Bay-Chi-all-90-00-gr},\ref{Bay-Hell-95-00-gr-kl-Bereich}
	show the result of a Bayesian peak detection procedure.   
	One knows that the lines in the spectrum are Gaussians 
	with a standard deviation $\sigma$ = 1.70 (1.59)\,keV 
	corresponding to 4.0 (3.7)\,keV FWHM.  
	Given the position of a line, 
	we used Bayes theorem to infer the contents of the line 
	and the level of a constant background. 
	We first describe the procedure summarily 
	and then give some mathematical details.

	Bayesian inference yields 
	the joint probability distribution for both parameters. 
	The background level was integrated out. This yielded 
	the distribution of the line contents. 
	If the distribution has its maximum at zero contents, 
	only an upper limit for the contents can be given 
	and the procedure does not suggest the existence of a line. 
	If the distribution has its maximum at non-zero contents, 
	the existence of a line is suggested and one can define 
	the probability K that there is a line with non-zero contents. 
	It is associated with an error interval containing 
	with probability K the true value of the contents.
	The Bayesian probability K agrees with the confidence of classical 
	statistics in the case of a Gaussian likelihood function.

	For every energy E of the spectra of Figs. 
\ref{Sum_spectr_Alldet_1990-2000},\ref{Sum_spectr_1-2-3-5det_95-2000},\ref{Sum_spectr_5det},
	we have determined the probability K that there is a line at E. 
	All the remainder of the spectrum was considered 
	to be background in this search. The result is given 
	on the left-hand sides of Figs. 
\ref{Bay-Alles-90-00-gr-kl-Bereich},\ref{Bay-Chi-all-90-00-gr},\ref{Bay-Hell-95-00-gr-kl-Bereich}.

	We define the Bayesian procedure in some more detail. 
	It starts from the distribution 
	$p(x_1...x_M|\rho,\eta)$
	of the count rates $x_1 ...x_M$ in the bins 1..M of the spectrum - 
	given two parameters $\rho,\eta$.
	The distribution $p$ is the product 
\beq{}
	p(x_1...x_M|\rho,\eta)~
	=~\prod_{k=1}^{M} \frac{{\lambda_k}^{x_k}}{x_k!}~e^{-\lambda_k}
\label{probab}
\eeq	
	of Poissonians for the individual bins. 
	The expectation value $\lambda_k$ is the superposition 
\beq{}
	\lambda_k~=~\rho~[\eta~ f_1(k) + (1 - \eta)~ f_2(k)]
\label{lambda-h}
\eeq
	of the form factors f$_1$ of the line 
	and f$_2$ of the background; 
	i.e. f$_1$(k) is the Gaussian centered 
	at E with the above-mentioned standard deviation
	and f$_2$(k)$\equiv$f$_2$ is a constant.

	Since
\beq{}
	\sum_{k=1}^{M} f_\nu(k)~=~1   \hspace{1.5cm}for~~ \nu=1,2,
\label{sum-f}
\eeq
	one has
\beq{}
	\sum_{k=1}^{M} \lambda_k~=~\rho. 
\label{sum-l}
\eeq
 
	Hence, $\rho$ parametrizes the total intensity in the spectrum, 
	and $\eta$ is the relative
	intensity in the Gaussian line.

	Prior distributions are given by Jeffreys' rule 
	($\S$5.35 of 
\cite{Hagan94}). 
	The parameter $\rho$ is integrated 
	out and the posterior distribution 
	P($\eta|x_1...x_M$)
	of the relative contents $\eta$ of the spectral line is obtained.

	The peak detection procedure yields lines 
	at the positions of known weak $\gamma$-lines 
	from the decay of $^{214}{Bi}$ at 2010.7, 2016.7, 2021.8 
	and 2052.9\,keV 
\cite{Tabl-Isot96}. 
	In addition, a line centered at 2039\,keV shows up. 
	This is compatible with the Q-value 
\cite{New-Q-2001,Old-Q-val}
	of the double beta decay process. We emphasize, 
	that at this energy no $\gamma$-line is expected 
	according to the compilations in 
\cite{Tabl-Isot96}. 
	We do not find 
	indications for the lines from $^{56}{Co}$ at 2034.7\,keV 
	and 2042\,keV 
	discussed earlier  
\cite{Diss-Dipl} 
	(Figs. 
\ref{Bay-Alles-90-00-gr-kl-Bereich},\ref{Bay-Chi-all-90-00-gr}). 

	Bayesian peak detection suggests a line at Q$_{\beta\beta}$ 
	whether or not one includes detector Nr. 4 
	without muon shield (Figs. 
\ref{Bay-Alles-90-00-gr-kl-Bereich},\ref{Bay-Chi-all-90-00-gr}). 
	The line is also suggested in Fig. 
\ref{Bay-Hell-95-00-gr-kl-Bereich}
	after removal of multiple site events (MSE), see below.


	On the left-hand side of Figs. 
\ref{Bay-Alles-90-00-gr-kl-Bereich},\ref{Bay-Chi-all-90-00-gr},\ref{Bay-Hell-95-00-gr-kl-Bereich},  
	the background intensity (1-$\eta$) 
	identified by the Bayesian procedure 
	is too high because the procedure averages the 
	background over all the spectrum (including lines) 
	except for the line it is trying to single out. 
	Therefore on the right-hand side of Figs. 
\ref{Bay-Alles-90-00-gr-kl-Bereich},\ref{Bay-Chi-all-90-00-gr},\ref{Bay-Hell-95-00-gr-kl-Bereich}, 
	the peak detection procedure is carried out within 
	an energy interval that does not contain 
	(according to the left-hand side) lines other 
	than the one at Q$_{\beta\beta}$. 
	Still the interval is broad enough 
	(about $\pm$ 5 standard deviations of 
	the Gaussian line) for a meaningful analysis. 
	We find the probability K = 96.5$\%$ that 
	there is a line at 2039.0\,keV in the spectrum shown in Fig. 
\ref{Sum_spectr_Alldet_1990-2000}.
	This is a confidence level of 2.1 $\sigma$ 
	in the usual language.  
	For the spectrum shown in Fig. 
\ref{Sum_spectr_1-2-3-5det_95-2000},  
	we find a probability for a line at 2039.0\,keV 
	of 97.4$\%$ (2.2 $\sigma$). 
	In this case
	the number of events is found to be 
	1.2 to 29.4 with 95$\%$ c.l..  
	It is 7.3 to 22.6 events with 68.3$\%$ c.l..
	The most probable number of events (best value) is  
	14.8 events.  


	We also applied the method recommended 
	by the Particle Data Group 
\cite{RPD98}.
	This method 
	(which does not use the information 
	that the line is Gaussian) finds 
	a line at 2039\,keV on a confidence level of 
	3.1 $\sigma$ (99.8$\%$ c.l.).  
	Such value is obtained in a wide range 
	of the assumed width of the signal 
	and the energy range of evaluation. 


\begin{table}[h]
\begin{center}
\newcommand{\m}{\hphantom{$-$}}
\renewcommand{\arraystretch}{1.}
\setlength\tabcolsep{7.7pt}
\begin{tabular}{c|c|c|c|c|c}
\hline
\hline
&&&&&\\
N	& N Bin	& Energy keV	& N	& N Bin	& Energy keV\\
\hline
&&&&&\\
1.	& 5653	& 2034.66	& 6.	& 5666	& 2039.33\\
2.	& 5658	& 2036.46	& 7.	& 5669	& 2040.41\\
3.	& 5660	& 2037.18	& 8.	& 5674	& 2042.21\\
4.	& 5664	& 2038.61	& 9.	& 5680	& 2044.37\\
5.	& 5665	& 2038.97	& 	&	&	\\
\hline
\hline
\end{tabular}
\end{center}
\caption{\label{List-Events}Events classified to be single 
	site events (SSE) by all three methods of PSA, 
	in the range 2034.1 - 2044.9\,keV ($\pm$ 3 $\sigma$ range around 
	Q$_{\beta\beta}$) of the spectrum taken 
	with enriched detectors Nr. 2,3,5 in the period 
	November 1995 - May 2000 (28.053\,kg\,y).
}
\end{table}


	From the analysis of the single site events, 
	we find after 28.053\,kg\,y of measurement 
	9 SSE events in the region 2034.1 - 2044.9\,keV 
	($\pm$ 3$\sigma$ around Q$_{\beta\beta}$)    
	(Table  
\ref{List-Events}).
	Analysis with the Bayesian method of the single site 
	event spectrum (Fig. 
\ref{Sum_spectr_5det}),  
	as described before, shows again evidence for a line   
	at the energy of Q$_{\beta\beta}$ (Fig. 
\ref{Bay-Hell-95-00-gr-kl-Bereich}). 
	Analyzing the range of 2032 - 2046\,keV, 
	we find 
	the probability of 96.8$\%$ 
	that there is a line at 2039.0\,keV.
	We thus see a signal of single site events,  
	as expected 
	for neutrinoless double beta decay, 
	precisely at the Q$_{\beta\beta}$ value obtained 
	in the precision experiment of 
\cite{New-Q-2001}. 	
	The analysis of the line at 2039.0\,keV 
	before correction for the efficiency yields 
	4.6\,events (best value) or 
	(0.3 - 8.0)\,events within 95$\%$ c.l. 
	((2.1 - 6.8)\,events within 68.3$\%$ c.l.). 
	Corrected for the efficiency to identify 
	an SSE signal by successive application of all 
	three PSA methods, which is 0.55 $\pm$ 0.10, 
	we obtain a \onbb signal with 92.4$\%$ c.l.. The signal is   
	(3.6 - 12.5)\,events 
	with 68.3 $\%$ c.l. (best value 
	8.3\,events). 

	The PDG method gives a signal at 2039.0\,keV of 2.8 $\sigma$ 
	(99.4$\%$).  
	The analysis given in Fig. 
\ref{Bay-Hell-95-00-gr-kl-Bereich}
	(left part) shows partly drastic differences 
	in the relative intensities of other identified lines 
	compared to the full spectra, 
	which should reflect their different composition of single site 
	and multiple site events.

	We emphasize that we find in all analyses 
	of our spectra a line at the value of Q$_{\beta\beta}$. 
	Under the assumption that the signal at Q$_{\beta\beta}$ is not 
	produced by a background line of presently unknown origin, 
	we can translate 
	the observed number of events into half-lives. 
	We give in Table 2 conservatively the values obtained with 
	the Bayesian method and not those obtained with the PDG method. 
	Also given in Table 2 are the effective neutrino masses    
	$\langle m \rangle $ deduced using the matrix elements of 
\cite{Sta90,Tom91}.

	We derive from the data taken with 46.502\,kg\,y 
	the half-life 
	${\rm T}_{1/2}^{0\nu} = (0.8 - 18.3) \times 10^{25}$ 
	${\rm y}$ (95$\%$ c.l.). 
	The analysis of the other data sets, shown in Table 
\ref{Results},
	and in particular of the single site events data, 
	which play an important role in our  
	conclusion, confirm this result.


\begin{table}[h]
\begin{center}
\newcommand{\m}{\hphantom{$-$}}
\renewcommand{\arraystretch}{1.}
\setlength\tabcolsep{5.7pt}
\begin{tabular}{c|c|c|c|c}
\hline
\hline
&&&&\\
Significan-	&	Detectors		
&	${\rm T}_{1/2}^{0\nu}	{\rm~ \,y}$	
& $\langle m \rangle $ eV	&	Conf. \\
	ce $[ kg\,y ]$	&&&& level\\
\hline
&&&&\\
	54.9813	
&	1,2,3,4,5	
&	$(0.80 - 35.07) \times 10^{25}$	
& (0.08 - 0.54)	
& 	$95\%  ~c.l.$	\\
	54.9813	
&	1,2,3,4,5
&	$(1.04 - 3.46) \times 10^{25}$	
& (0.26 - 0.47)	
& $68\%  ~c.l.$	\\
 	54.9813	
&	1,2,3,4,5
&	$1.61 \times 10^{25}$	
& 0.38 	
& Best Value	\\
\hline
 	46.502	
&	1,2,3,5
&	$(0.75 - 18.33) \times 10^{25}$	
& (0.11 - 0.56)
& $95\%  ~c.l.$	\\
	46.502	
&	1,2,3,5
&	$(0.98 - 3.05) \times 10^{25}$	
& (0.28 - 0.49)
& $68\% ~c.l.$	\\
	46.502	
&	1,2,3,5
&	$1.50 \times 10^{25}$	
& 0.39 
& Best Value		\\
\hline
	28.053	
&	2,3,5  SSE	
&	$(0.88 - 22.38) \times 10^{25}$	
& (0.10 - 0.51)	
& $90\% ~c.l.$		\\
	28.053	
&	2,3,5  SSE
&	$(1.07 - 3.69) \times 10^{25}$	
& (0.25 - 0.47)		
& $68\% ~c.l.$		\\
	28.053	
&	2,3,5 SSE
&	 $1.61 \times 10^{25}$	
& 0.38
& Best Value	\\
\hline
\hline
\end{tabular}
\end{center}
\caption[]{\label{Results}Half-life for the neutrinoless decay mode 
	and deduced effective neutrino mass 
	from the HEIDELBERG-MOSCOW experiment.
}
\end{table}


	The result obtained is consistent with the limits 
	given earlier by the HEI\-DELBERG-MOSCOW experiment 
\cite{HDM01}, 
	considering that the background had been determined 
	more conservatively there.
	It is also consistent with all other double beta experiments -  
	which still reach less sensitivity. 
	A second Ge-expe\-riment, 
	which has stopped operation in 1999 after 
	reaching a significance of 9\,kg\,y
	yields (if one believes their method of 'visual inspection' 
	in their data analysis) in a conservative analysis a limit of 
	${\rm T}_{1/2}^{0\nu} > 0.55 \times 10^{25}
	{\rm~ y}$  (90\% c.l.). 
	The $^{128}{Te}$ geochemical experiment 
	yields $\langle m_\nu \rangle < 1.1$\,eV (68$\%$ c.l.), 
	the $^{130}{Te}$ cryogenic experiment yields 
	$\langle m_\nu \rangle < 1.8$\,eV 
	and the CdWO$_4$ experiment $\langle m_\nu \rangle < 2.6$\,eV,  
	all derived with the matrix elements of 
\cite{Sta90,Tom91} 
	to make the results comparable to the present value 
	(for references see 
\cite{KK60Y}).


	Concluding we obtain, with more than 95$\%$ probability, 
	first evidence for the neutrinoless 
	double beta decay mode.


\begin{figure}[h]
\vspace{9pt}
\centering{
\includegraphics*[scale=0.43]
{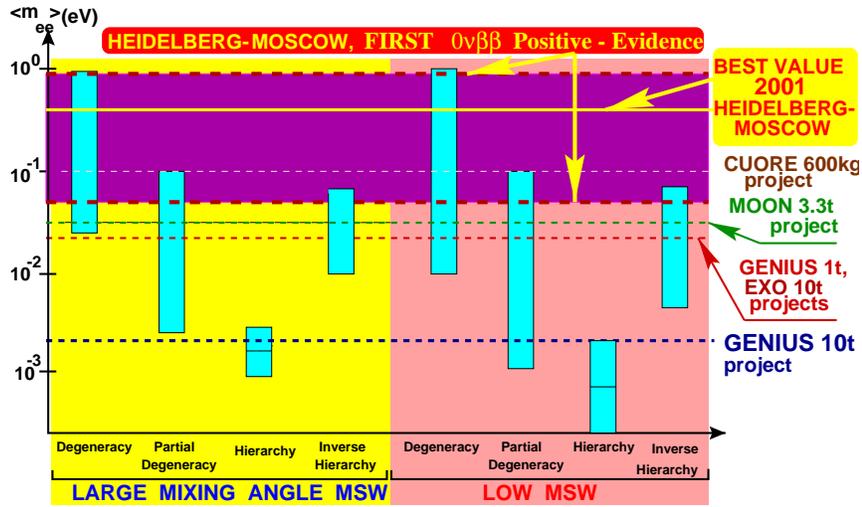}}
\caption[]{
	{The impact of the evidence obtained for neutrinoless 
	double beta decay in this paper (best value 
	of the effective neutrino mass 
	$\langle m \rangle$ = 0.39\,eV, 95$\%$ 
	confidence range (0.05 - 0.84)\,eV - 
	allowing already for an uncertainty of the nuclear 
	matrix element of a factor of $\pm$ 50$\%$ 
	on possible neutrino mass schemes. 
	The bars denote allowed ranges of $\langle m \rangle$ 
	in different neutrino mass scenarios, 
	still allowed by neutrino oscillation experiments (see 
\cite{KKPS1}). 
	Hierarchical models are excluded by the 
	new \onbb decay result. Also shown are 
	the expected sensitivities 
	for the future potential double beta experiments 
	CUORE, MOON, EXO  
	and the 1 ton and 10 ton project of GENIUS 
\cite{KK-00-NOON-NOW-NANP-Bey97-GEN-prop}.}
\label{Jahr00-Sum-difSchemNeutr}}
\end{figure}

	
	As a consequence, on the same confidence level, 
	lepton number is not conserved. 
	Further the neutrino is a Majorana particle. 
	We conclude from the various 
	analyses given above the effective mass 
	$\langle m \rangle $ 
	to be $\langle m \rangle $ 
	= (0.11 - 0.56)\,eV (95$\%$ c.l.), 
	with best value of 0.39\,eV. 
	Allowing conservatively for an uncertainty of the nuclear 
	matrix elements of $\pm$ 50$\%$
	(for detailed discussions of the status 
	of nuclear matrix elements we refer to 
\cite{Tom91,Vog2001,KK60Y,FaesSimc}) 
	this range may widen to 
	$\langle m \rangle $ 
	= (0.05 - 0.84)\,eV (95$\%$ c.l.). 

	In this conclusion, it is assumed that contributions 
	to \onbb decay from processes other than  
	the exchange of 
	a Majorana neutrino (see, e.g. 
\cite{Moh91,KK60Y}) 
	are negligible.

	With the limit deduced for the effective neutrino mass,  
	the HEIDELBERG-MOSCOW experiment excludes several 
	of the neutrino mass 
	scenarios 
	allowed from present neutrino oscillation experiments
	(see Fig.
\ref{Jahr00-Sum-difSchemNeutr}) 
	- allowing mainly only for degenerate 
	and partially degenerate mass 
	scenarios and an inverse hierarchy 3$\nu$ - scenario
	(the latter being, however, strongly disfavored 
	by a recent analysis
	of SN1987A.   
	In particular hierarchical mass schemes 
	are excluded at the present level of accuracy.

	Assuming the degenerate scenarios to be realized in nature 
	we fix - according to the formulae derived in 
\cite{KKPS1} - 
	the common mass eigenvalue of the degenerate neutrinos 
	to m = (0.05 - 3.4)\,eV. 
	Part of the upper range is already excluded by 
	tritium experiments, which give a limit of m $<$ 2.2\,eV (95$\%$ c.l.) 
\cite{Trit00}.
	The full range can only  partly 
	(down to $\sim$ 0.5\,eV) be checked by future  
	tritium decay experiments,  
	but could be checked by some future $\beta\beta$ 
	experiments (see, e.g. 
\cite{KK60Y,KK-00-NOON-NOW-NANP-Bey97-GEN-prop}).
	The deduced best value for the mass 
	is consistent with expectations from experimental 
	$\mu ~\to~ e\gamma$
	branching limits in models assuming the generating 
	mechanism for the neutrino mass to be also responsible 
	for the recent indication for as anomalous magnetic moment 
	of the muon
\cite{MaRaid01}.
	It lies in a range of interest also for Z-burst models recently 
	discussed as explanation for super-high energy cosmic ray events 
	beyond the GKZ-cutoff 
\cite{PW01-Wail99}.
	
	The neutrino mass deduced allows neutrinos to still play 
	an important role as hot dark matter in the Universe.

	New approaches and considerably enlarged experiments 
	(as discussed, e.g. in 
\cite{KK-00-NOON-NOW-NANP-Bey97-GEN-prop})
	will be required in future 
	to fix the neutrino mass with higher accuracy.

\vspace{.5cm}
{\large\bf Acknowledgments}

\vspace{.5cm}
\noindent
	One of the authors - H.V. Klapdor-Kleingrothaus - 
	would like to 
	thank all colleagues who have contributed to the 
	experiment over the last decade of operation. 
	They are particularly grateful to Dr. F. Petry, Dr. B. Maier, 
	Dr. J. Hellmig and Dr. B. Majorovits who have developed 
	the pulse shape discrimination methods and set up the VME electronics 
	applied in the experiment since 1995. 
	They thank Dr. T. Kihm for his highly efficient permanent 
	support in the field of electronics and Dr. G. Heusser for his 
	important contribution in the field of low-level techniques. 
	They thank Mr. H. Strecker for his invaluable technical 
	contributions.  
	They thank Prof. D. Schwalm for his scientific interest 
	and his efficient support of this project and would like 
	to thank the director of the Gran Sasso Underground Laboratory, 
	Prof. E. Bettini and also the former directors 
	of Gran Sasso Profs. P. Monacelli and E. Bellotti, 
	for invaluable support. 
	Our thanks extend also to the technical staff of the 
	Max-Planck Institut f\"ur Kernphysik and 
	of the Gran Sasso Underground Laboratory. 
	We thank Perkin Elmer (former ORTEC) Company, 
	and in particular Dr. M. Martini, and Dr. R. Collatz 
	for the fruitful cooperation.
	We are grateful to Dr. G. Sawitzki and Prof. W. Beiglb\"ock from 
	the Institute for Applied Mathematics of the University 
	of Heidelberg for encouraging discussions. 
	We acknowledge the invaluable support from BMBF 
	and DFG of this project.
	We are grateful to the former State Committee of Atomic 
	Energy of the USSR for providing the enriched material 
	used in this experiment.


\end{document}